\documentclass{aa}
\pdfoutput=1
\usepackage{graphicx}
\usepackage{txfonts}
\usepackage{natbib}
\bibpunct{(}{)}{;}{a}{}{,}
\usepackage{amssymb}

\hypersetup{pdfauthor={Magnus Vilhelm Persson},
			pdftitle={The deuterium fractionation of water on solar-system scales in deeply-embedded low-mass protostars}}

\begin{document}
%
\newcommand{\figref}[1]{Fig.~\ref{#1}} 
\newcommand{\up}[1]{$^{#1}$}
\newcommand{\down}[1]{$_{#1}$}
\newcommand{\tblmark}[1]{\tablefootmark{#1}}
\newcommand{\tabref}[1]{Table~\ref{#1}}
\newcommand{\spitzer}{\textit{Spitzer}}
\newcommand{\herschel}{\textit{Herschel}}
\newcommand{\tento}[1]{$\times10^{#1}$}
\newcommand{\tsim}{$\sim$}
\newcommand{\kms}{km~s$^{-1}$}
\newcommand{\mjykms}{mJy~\kms}
\newcommand{\jykms}{Jy~km~s$^{\rm-1}$}
\newcommand{\Msun}{$M_{\rm\sun}$}
\newcommand{\Lsun}{$L_{\rm\sun}$}
\newcommand{\Mearth}{$M_{\rm\oplus}$}
\newcommand{\tlambda}{$\lambda$} 
\newcommand{\farcmRA}{$\hbox{$.\!\!^{\mathrm{s}}$}$} 
\newcommand{\tfarcmRA}{\hbox{$.\!\!^{\mathrm{s}}$}} 
\newcommand{\arcm}{^{\prime}}			
\newcommand{\tarcm}{$^{\prime}$} 		
\newcommand{\eu}{$E_\mathrm{u}$} 		
\newcommand{\dlq}[1]{``}
\newcommand{\drq}[1]{''}
\newcommand{\isowater}{{H$_2^{18}$O}}
\newcommand{\normalwater}{H$_2$O}
\newcommand{\dimethylether}{CH$_3$OCH$_3$}
\newcommand{\ethylcyanide}{C$_2$H$_5$CN}
\newcommand{\sulfurdioxide}{SO$_2$}
\newcommand{\trans}[6]{$#1_{#2,#3}-#4_{#5,#6}$}

   \title{The deuterium fractionation of water on solar-system scales in deeply-embedded low-mass protostars\thanks{Based on observations carried out with the IRAM Plateau de Bure
   Interferometer.}}

   \subtitle{}
   \author{M.~V. Persson
          \inst{1}\fnmsep\inst{2}\fnmsep\inst{3}
          \and
          J.~K. J{\o}rgensen\inst{2}\fnmsep\inst{1}	
          \and E.~F. van Dishoeck
          \inst{3}\fnmsep\inst{4}
	\and D. Harsono\inst{3}
          }

   \institute{Centre for Star and Planet Formation,
   Natural History Museum of Denmark,
   University of Copenhagen,
   {\O}ster Voldgade 5-7, \\DK-1350,
  Copenhagen K, Denmark\\
              \email{magnusp@strw.leidenuniv.nl}
         \and
            Niels Bohr Institute, University of Copenhagen, Juliane
            Maries Vej 30, DK-2100 Copenhagen {\O}, Denmark
         \and
             Leiden Observatory, Leiden University, P.O. Box 9513, NL-2300 RA Leiden, The Netherlands
        \and
            Max-Planck Institute f\"{u}r extraterrestrische Physik (MPE), Giessenbachstrasse, 85748 Garching, Germany
             }

   \date{Received , 2013; accepted , 2014}

 
  \abstract
	{The chemical evolution of water through the star formation
          process directly affects the initial conditions of planet
          formation. The water deuterium fractionation (HDO/H$_2$O
          abundance ratio) has traditionally been used to infer the
          amount of water brought to Earth by comets. Measuring this
          ratio in deeply-embedded low-mass protostars makes it
          possible to probe the critical stage when water is transported
          from clouds to disks in which icy bodies are formed.}
	{We aim to determine the HDO/H$_2$O abundance ratio in the warm gas in the inner 150 AU for three deeply-embedded low-mass protostars NGC~1333-IRAS~2A, IRAS~4A-NW, and IRAS~4B through high-resolution interferometric observations of isotopologues of water.}
   	{We present sub-arcsecond resolution observations of the
          $3_{1,2}-2_{2,1}$ transition of HDO at 225.89672~GHz in
          combination with previous observations of the
          $3_{1,3}-2_{2,0}$ transition of H$_2^{18}$O at 203.40752~GHz
          from the Plateau de Bure Interferometer toward three
          low-mass protostars. The observations have similar angular
          resolution (0\farcs7 - 1\farcs3), probing scales
          $R\lesssim150$~AU. In addition, observations of the
          $2_{1,1}-2_{1,2}$ transition of HDO at 241.561~GHz toward
          IRAS~2A are presented to constrain the excitation
          temperature. A direct and model independent
          HDO/H$_2$O abundance ratio is determined for each source and
          compared with HDO/H$_2$O ratios derived from spherically
          symmetric full radiative transfer models for two sources. }
   	{From the two HDO lines observed toward IRAS~2A, the
          excitation temperature is found to be
          $T_\mathrm{ex}=124\pm60$~K. Assuming a similar excitation
          temperature for H$_2^{18}$O and all sources, the HDO/H$_2$O
          ratio is $7.4\pm2.1\times10^{-4}$ for IRAS~2A,
          $19.1\pm5.4\times10^{-4}$ for IRAS~4A-NW, and
          $5.9\pm1.7\times10^{-4}$ for IRAS~4B. The abundance ratios
          show only a weak dependence on the adopted excitation
          temperature. The
          abundances derived from the radiative transfer models
          agree with the direct determination of the HDO/H$_2$O
          abundance ratio for IRAS~16293-2422 within a factor of 2-3, and for IRAS~2A within a factor of 4; the difference is mainly due to optical depth effects in the HDO line.}
 	{Our HDO/H$_2$O ratios for the inner regions (where $T>100$~K)
          of four young protostars are only a factor
          of 2 higher than those found for pristine, solar system
          comets. These small differences suggest that little
          processing of water occurs between the deeply embedded stage
          and the formation of planetesimals and comets. 
 	}
   \keywords{astrochemistry -- ISM: abundances -- stars: formation -- stars: individual objects: NGC~1333 IRAS~2A, NGC~1333 IRAS~4A, NGC~1333 IRAS~4B}

   \maketitle


\section{Introduction}

Water is a very important molecule through all stages of star and
planet formation. In dense clouds, it is one of the dominant
reservoirs of oxygen, both as a gas at high temperatures or as ice in grain mantles in cold regions. It also aids in the sticking of grains outside the snow line, and is crucial for the emergence of life on Earth. This makes it vital to understand the chemical evolution of
water during the different stages of star formation, and to identify
the mechanism by which it was brought to the Earth and Earth-like planets.

One way to gain insight into the chemical evolution of water is to
compare the level of deuterium fractionation in water (the HDO/H$_2$O
ratio) at various stages of star and planet formation. In our own
solar system, several measurements of the HDO/H$_2$O
ratio\footnote{$\mathrm{D/H}=0.5\times\mathrm{HDO/H_2O}$} exist for
long-period comets from the Oort cloud \citep[][and references
therein]{mumma11}. Up until recently the mean HDO/H$_2$O ratio for
these objects was $6.4\pm1.0\times 10^{-4}$
\citep{villanueva09,jehin09}, significantly above the value for the Earth's oceans $3.114\pm 0.002\times 10^{-4}$
\citep[VSMOW\footnote{Vienna Standard Mean Ocean Water},][and
ref. therein]{delaeter03}. Both values are well above the protosolar
nebula value of 2$\times$D/H=$0.42\pm0.08\times10^{-4}$
\citep{geiss98,lellouch01} indicating that deuterium fractionation of
water has taken place. 
Because of the observed difference between the Earth and comets, it has been argued that only a small fraction ($\le10\%$) of the Earth's water could have its origin in comets \citep{morbidelli00,drake05}.

The Jupiter family comet Hartley~2 observed with the \herschel\ Space
Observatory was found to have a HDO/H$_2$O ratio of
$3.2\pm0.5\times10^{-4}$, significantly lower than the value for Oort
cloud comets, and comparable to VSMOW \citep{hartogh11}. Jupiter family comets have relatively short periods ($\lesssim20$~years), with orbits more or less confined to the ecliptic plane, and originate in the Kuiper belt \citep{duncan04,mumma11}. Oort cloud comets have longer periods, orbits distributed isotropically, and originate in the Oort cloud. In 2012 \citeauthor{bockelee12} observed the Oort cloud comet Garradd with \herschel\ and deduced a HDO/H$_2$O ratio of $4.12\pm0.44\times10^{-4}$. This is significantly lower than
previous measurements of Oort cloud comets. Recently, \citet{lis13}
measured the HDO/H$_2$O ratio in the Jupiter family comet
Honda-Mrkos-Pajdusakov (HMP) using \herschel\ to
$<4.0\times10^{-4}$. These latest measurements indicate a certain
range in the water deuterium fractionation in comets, possibly
dependent on the precise formation location and subsequent migration,
in which Jupiter family comets no longer have a distinctly lower
HDO/H$_2$O ratio than Oort cloud comets. Simulations suggest that the
deuterium fractionation in the protosolar nebula disk increases with
distance from the forming sun \citep{hersant01,kavelaars11} -- but
whether the observed ratios are consistent with these simulations is
still unclear.

Measurements of the water deuterium fractionation toward protostars
provide a different perspective. By determining the HDO/H$_2$O ratio
of material that enters protoplanetary disks, it is possible to set
the initial level of deuterium fractionation in comet forming zones as
input for any subsequent evolutionary models. So far, determinations
of HDO/H$_2$O have given a wide range of results. \citet{parise03}
used ground-based infrared observations of the stretching bands of OH
and OD in water ice in the cold outer parts of protostellar envelopes
and found \textit{upper limits} ranging from 0.5\% to 2\% for the
HDO/H$_2$O ratios. Studies using gaseous HDO lines differ in the
interpretation, with cold HDO/H$_2$O ratios derived from single-dish
data ranging from cometary values
\citep[$2\times10^{-4}$,][]{stark04}, to a few $\times10^{-2}$
\citep{parise05, liu11, coutens12}.  In the inner warm ($T>100$ K)
region, HDO/H$_2$O ratios $\geq$1\% have also been inferred based on
analysis of many HDO and water lines ($E_\mathrm{u}$ up to 450~K)
\citep{liu11,coutens12}. In contrast, \citet{jorgensen10b} derived a
HDO/H$_2$O ratio toward the Class~0 protostar NGC~1333-IRAS~4B of
$<6.4\times10^{-4}$ from the interferometric Submillimeter Array (SMA) and
Plateau de Bure (PdBI) observations, up to two orders of magnitude
lower.

There are two main issues with determining the water abundances in the
warm gas in the inner few hundred au. First, the relatively large
beam sizes ($>10$\arcsec) of single-dish telescopes are more sensitive
to extended emission than the small $\sim 1\arcsec$ scales on which
the warm water is found. Second, many observed water lines suffer from
high optical depths, even for isotopologues, making it difficult to
infer reliable column densities and abundances for H$_2$O in the warm
($>100$~K) gas. Indeed, \citet{visser13} find that the
$3_{1,2}-3_{0,3}$ lines ($E_u=249$~K) of H$_2^{16}$O, H$_2^{18}$O and
even H$_2^{17}$O observed by the HIFI instrument \citep{graauw10}
aboard \herschel\ probably show optically thick emission on
$R\sim100$~AU scales in the well-studied Class 0 protostar
NGC1333-IRAS~2A. Taking the optical depth properly into account lowers
the HDO/H$_2$O ratio for IRAS~2A from $\ge0.01$ to $1\times10^{-3}$
\citep{visser13}. 

Recent high-resolution interferometric ground-based observations have
made it possible to estimate the abundances of both H$_2$O and HDO in
the warm gas on small scales. \citet{jorgensen10a} targeted the
H$_2^{18}$O $3_{1,3}-2_{2,0}$ line at 203 GHz with PdBI, which has a
much lower Einstein $A$ coefficient than the lines observed with
\herschel\ and thus suffers less from optical depth effects. Using
SMA data of the HDO $3_{1,2}-2_{2,1}$ transition at 225 GHz, the above mentioned
HDO/H$_2$O limit of $<6.4\times10^{-4}$ was found toward
NGC~1333-IRAS~4B \citep{jorgensen10b}.
\citet{coutens13} obtained values of $4 - 30\times10^{-4}$ for
IRAS~4A and $1 - 37\times10^{-4}$ for IRAS~4B using new \herschel\
data combined with the H$_2^{18}$O column densities found by
\citet{persson12} based on interferometric data.

\citet{taquet13b} derived the warm HDO/H$_2$O ratios for IRAS~2A and
IRAS~4A using the determinations of the H$_2^{18}$O column density from
\citet{persson12} and lower spatial- and spectral-resolution
interferometric observations targeting the HDO $4_{2,2}-4_{2,3}$
transition (143.7~GHz, $E_\mathrm{u}=316.2$~K) together with single
dish \citep{liu11} observations of HDO.  Using non-local
thermal equilibrium (non-LTE) large velocity gradient (LVG) analysis,
ratios in the range of $30-800\times10^{-4}$ for IRAS~2A and
$50-300\times10^{-4}$ for IRAS~4A were obtained, depending on the
assumed density, with the high-density
($n_\mathrm{H_2}=1\times10^8$~cm$^{-3}$) solution giving the lowest
ratios.  Finally, using Atacama Large Millimeter/submillimeter Array
(ALMA) and SMA observations, \citet{persson13a} estimated the
HDO/H$_2$O ratio of the warm water toward IRAS~16293-2422~A to
$9.2\pm2.6\times10^{-4}$. 

Most of these new values, summarized in Table~A.1, are significantly lower than the previous estimates of a few percentage points, indicating that the deuterium fractionation in
warm water may not be as enhanced as previously thought.

In order to settle the question of the HDO/H$_2$O abundance ratios
in warm gas, we have obtained high spatial resolution interferometric
observations of HDO to complement previous high quality data of
H$_2^{18}$O for three sources. The observations are sensitive to
scales comparable to disks ($\sim 1\farcs3$, $R \lesssim150$ AU),
allowing us to determine the HDO/H$_2$O ratio in the innermost parts
of young protostars where the disk is forming.

This paper presents observations of HDO \trans{3}{1}{2}{2}{2}{1}\ at
225.9~GHz toward the three well-known deeply-embedded low-mass
protostars in the NGC~1333 region of the Perseus molecular cloud
(\citealt{jennings87,sandell91} --- IRAS~2A, IRAS~4A, and IRAS~4B in
the following) at a distance of 250~pc \citep{enoch06}. The three
sources are ideal to compare because of their similar masses,
luminosities, and location \citep{jorgensen06}. \citet{persson12} presented observations of the H$_2^{18}$O \trans{3}{1}{3}{2}{2}{0}\ transition at 203.4~GHz toward the same sources. With these
two datasets we can deduce a HDO/H$_2$O ratio in the warm gas in the
inner 150~AU radius using the same method and similar sensitivity and
resolution as for IRAS~16293-2422, bringing the total number of
low-mass protostars for which high quality data exist to four. In
addition, observations of HDO $2_{1,1}-2_{1,2}$ at 241.6~GHz toward
IRAS~2A were acquired to empirically constrain the excitation
temperature.  The paper is laid out as follows.
Section~\ref{section:observations} describes the details of the various observations. Sections~\ref{section:results} and \ref{section:analysis}
present the results and analysis -- including intensities, column
densities, spectra, maps, and line radiative transfer modeling of the
observed emission. In Sect.~\ref{section:discussion} we discuss the
deduced HDO/H$_2$O ratio and compare it to other studies of protostars
and with solar system objects.


\section{Observations}\label{section:observations}

Three low-mass protostars, IRAS~2A, IRAS~4A, and IRAS~4B in the embedded cluster NGC~1333 were observed using the PdBI. All sources were observed with two different receiver setups. One setup was tuned to the \isowater\ \trans{3}{1}{3}{2}{2}{0}\ transition at 203.40752~GHz, for details about those observations see \citet{jorgensen10a} and \citet{persson12}. In the other setup, the receivers were tuned to the HDO \trans{3}{1}{2}{2}{2}{1}\ transition at 225.89672~GHz.
IRAS~2A was observed in the C configuration on 27 and 28 November 2011 (8~hours; including calibration) and the B configuration on 12 March 2012 (2~hours). IRAS~4A and IRAS~4B were observed in track sharing mode in the B configuration on 12 March 2012 (4~hours), and C on 15, 27, and 21 March and on 2 April (3~hours). For IRAS~2A the combined dataset covers baselines from 15.8 to 452.0~m (11.9 to 340.5~k\tlambda), for IRAS~4A from  15.0 to 452.0~m (11.3 to 340.5~k\tlambda), and for IRAS~4B from 16.3 to 451.9~m (12.3 to 340.5~k\tlambda). The correlators were set up with one unit with a bandwidth of 40~MHz (47.6~\kms) centered on the frequency of the HDO line (225.89672~GHz), providing a spectral resolution on 460 channels of 0.087~MHz (0.104~\kms) width\footnote{Spectra and maps are available through \url{http://vilhelm.nu}.}.

The data were calibrated and imaged using the CLIC and MAPPING packages, part of the IRAM GILDAS software. Regular observations of the nearby, strong quasars 0333+321, 3C84, and 2200+420 or 1749+096 were used to calibrate the complex gains and bandpass, while MWC349 and 0333+321 were observed to calibrate the absolute flux scale. During the calibration procedure integrations with significantly deviating amplitude and or phases were flagged and the continuum was subtracted before Fourier transformation of the line data.

In addition to these observations, the HDO $2_{1,1}-2_{1,2}$ line at 241.56155~GHz was observed toward IRAS~2A on 7 and 17 November 2012 in the C configuration for 8~hours covering baselines from 15.5 to 175.8~m (12.5 to 141.7~k$\lambda$). The correlators were set up with one unit covering 80~MHz (89.0~\kms) with a spectral resolution of 0.174~MHz (0.1939~\kms) on 460 channels. The same calibration steps as described above were followed.

The resulting beam sizes using natural weighting and other parameters for the observations are given in \tabref{table:observations}. The field of view is roughly 22\arcsec\ (HPBW) at 225.9~GHz and the continuum sensitivity is limited by the dynamical range of the interferometer.
The uncertainty in fluxes is dominated by the accuracy of the flux calibration, typically about 20\%.

   \begin{table}
      \caption[]{Parameters for the observations tuned to HDO at 225.9~GHz and 241.6~GHz.}
         \label{table:observations}
     $$ 
         \begin{array}{p{0.3\linewidth}lll}
            \hline\hline
            \noalign{\smallskip}
            Source              &  \mathrm{Beam}\tblmark{a} & \mathrm{PA} &  \mathrm{RMS} \\
                          		&  \mathrm{[^{\prime\prime}]} & \mathrm{[^\circ]} &  \mathrm{[\tblmark{b}]} \\
            \noalign{\smallskip}
            \hline
            \noalign{\smallskip}
            IRAS~2A	            & 1.3\times 1.0	& 23.3	& 16.0   \\
            IRAS~2A\tblmark{c}	& 1.2\times 0.9	& 13.3	& 10.5   \\
            IRAS~4A	            & 1.1\times 0.8	& 11.3	& 12.5   \\
            IRAS~4B	            & 1.1\times 0.8	& 10.6	& 10.2   \\
            \noalign{\smallskip}
            \hline
         \end{array}
     $$ 
         \tablefoottext{a}{Major $\times$ minor axis.}
         \tablefoottext{b}{Units of mJy~beam$^{-1}$~channel$^{-1}$.}
         \tablefoottext{c}{For the 241.6~GHz observations.}
   \end{table}


\section{Results}\label{section:results}

\begin{figure}[ht]
   \centering
    \includegraphics{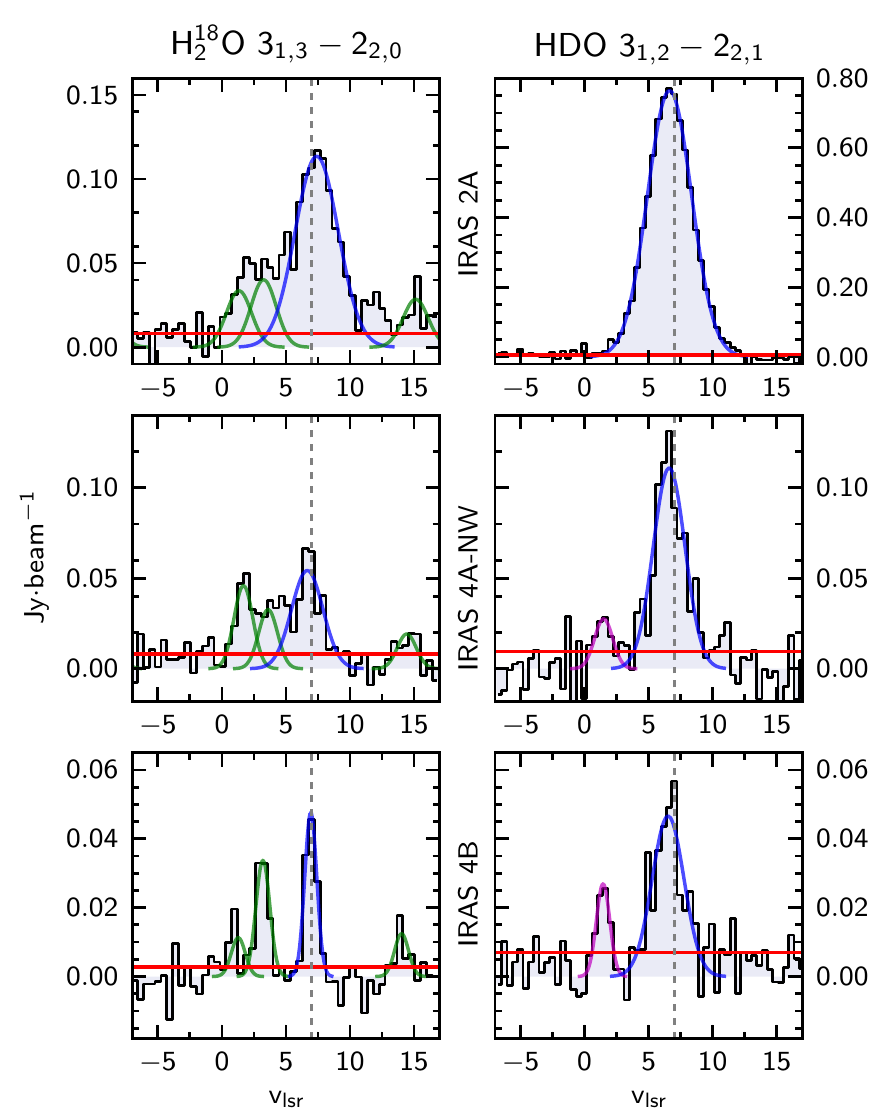}
      \caption{Continuum subtracted spectra of the H$_2^{18}$O $3_{1,3}-2_{2,0}$ and HDO $3_{1,2}-2_{2,1}$ lines toward the three sources NGC-1333~IRAS~2A, IRAS~4A-NW, and IRAS~4B. The H$_2^{18}$O spectra are from \citet{jorgensen10a} and \citet{persson12}. The green lines shows the CH$_3$OCH$_3$, and blue H$_2^{18}$O. In some of the HDO spectra (bottom two), a second line can be seen (magenta) at $~1.5$~\kms, this is from CH$_3$OCHO. The parental cloud velocity is shown by the dotted line ($v_\mathrm{lsr}=7$~\kms) and the Gaussian fits (green, magenta, and blue) to the lines are plotted along with the RMS (red). We note the different scales of the spectra toward IRAS~2A.
              }
         \label{figure:spectra_all}
       \includegraphics{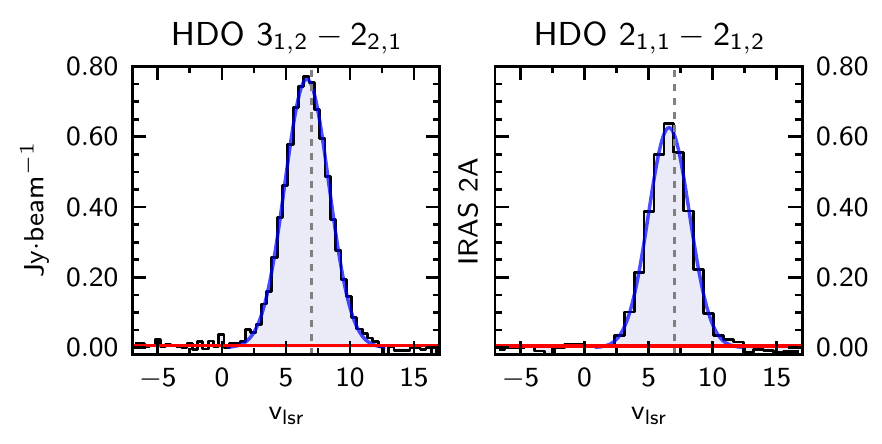}
         \caption{Continuum subtracted spectra of the HDO $3_{1,2}-2_{2,1}$ (225 GHz) and $2_{1,1}-2_{1,2}$ (241 GHz) lines toward IRAS2A. The parental cloud velocity is shown by the dotted line ($v_\mathrm{lsr}=7$~\kms) and the Gaussian fits (blue) to the HDO lines are plotted along with the RMS (red).
                 }
            \label{figure:spectra_i2a}
      \end{figure}

Both the targeted lines and the continuum are detected with high
signal to noise in all three sources (Fig.~\ref{figure:spectra_all}).
Elliptical Gaussian fits to the (\textit{u,v})-plane of the
continuum are tabulated in Table~\ref{table:continuum}. The continuum
emission from the sources agrees with previous observations
\citep[e.g.,][]{jorgensen07,jorgensen10a,persson12}, following a
power-law ($F_\nu\propto\nu^\alpha$) with exponents expected from
thermal dust continuum emission (i.e., $\alpha\sim 2-3$). This shows
that the flux calibrations for the different observations are accurate
to within roughly $20\%$.

   \begin{table}
      \caption[]{Parameters of the continuum determined from elliptical Gaussian fits in the (\textit{u,v})-plane.}
         \label{table:continuum}
     $$ 
         \begin{array}{p{0.3\linewidth}lll}
            \hline\hline
            \noalign{\smallskip}
            Source              &  \mathrm{Intensity} & \mathrm{PA} &  \mathrm{Size} \\
                                &  \mathrm{[Jy]} &    \mathrm{[^\circ]}      &  \mathrm{[^{\prime\prime}]} \\
            \noalign{\smallskip}
            \hline
            \noalign{\smallskip}
            IRAS~2A		        & 0.4 & 27.5& 1.1 \times 10 \\
            IRAS~2A\tblmark{a}   & 0.5 & 23.7& 1.3 \times 10 \\
            IRAS~4A-NW	        & 1.7 & - 	& 2.2 \\
            IRAS~4A-SE	        & 2.1 & - 	& 1.6 \\
            IRAS~4B		        & 1.1 & 6.7 & 2.2 \times 1.3 \\
            \noalign{\smallskip}
            \hline
         \end{array}
     $$ 
         \tablefoottext{a}{For the 241.6~GHz observations.}
   \end{table}

To identify the lines, the Jet Propulsion Laboratory \citep[JPL, ][]{pickett98} and the Cologne Database for Molecular Spectroscopy \cite[CDMS, ][]{muller01} line-lists were queried through the Splatalogue\footnote{\url{http://splatalogue.net}} interface. The molecular parameters for the lines are tabulated in Table~\ref{table:molecules}. The continuum subtracted spectra of both the H$_2^{18}$O $3_{1,3}-2_{2,0}$ and the HDO $3_{1,2}-2_{2,1}$ transition extracted toward the continuum peak for all the sources (excluding IRAS~4A-SE) are shown in \figref{figure:spectra_all} \citep[H$_2^{18}$O spectra from ][]{jorgensen10a,persson12}. Fig.~\ref{figure:spectra_i2a} shows the spectra for the HDO $3_{1,2}-2_{2,1}$ and $2_{1,1}-2_{1,2}$ transitions toward IRAS~2A. All lines used in this analysis are detected at high signal-to-noise (i.e.\ $\gtrsim10\sigma$). In the H$_2^{18}$O spectra, the lines seen besides water are due to dimethyl ether (CH$_3$OCH$_3$), for more details and information about the line assignment see \citet{jorgensen10a,persson12}. In the new HDO spectra for the $3_{1,2}-2_{2,1}$ transition, at $v_\mathrm{lsr}\approx 1.5$~\kms\ a line from methyl formate (CH$_3$OCHO, $E_\mathrm{u}=36.3$~K) is seen toward IRAS~4A and IRAS~4B (Fig.~\ref{figure:spectra_all}). The targeted HDO line is strong in all the sources. Just as with the H$_2^{18}$O line, no HDO is detected toward IRAS~4A-SE. 

As noted in \citet{jorgensen10a} and \citet{persson12,persson13a} the
H$_2^{18}$O lines are not found to be masing in these sources on the
observed scales. This is in contrast to the $3_{1,3}-2_{2,0}$
transition of the main isotopologue, H$_2^{16}$O, which is masing in many
sources on single-dish scales \citep{cernicharo94}. 
The narrow widths of all detected H$_2^{18}$O and HDO lines presented
here demonstrate that the lines do not originate from any outflow in
these sources, where the lines have been observed to have broader
widths \citep[i.e., $>5$~\kms][]{kristensen10}.

   \begin{figure}[ht]
   \centering
    \includegraphics{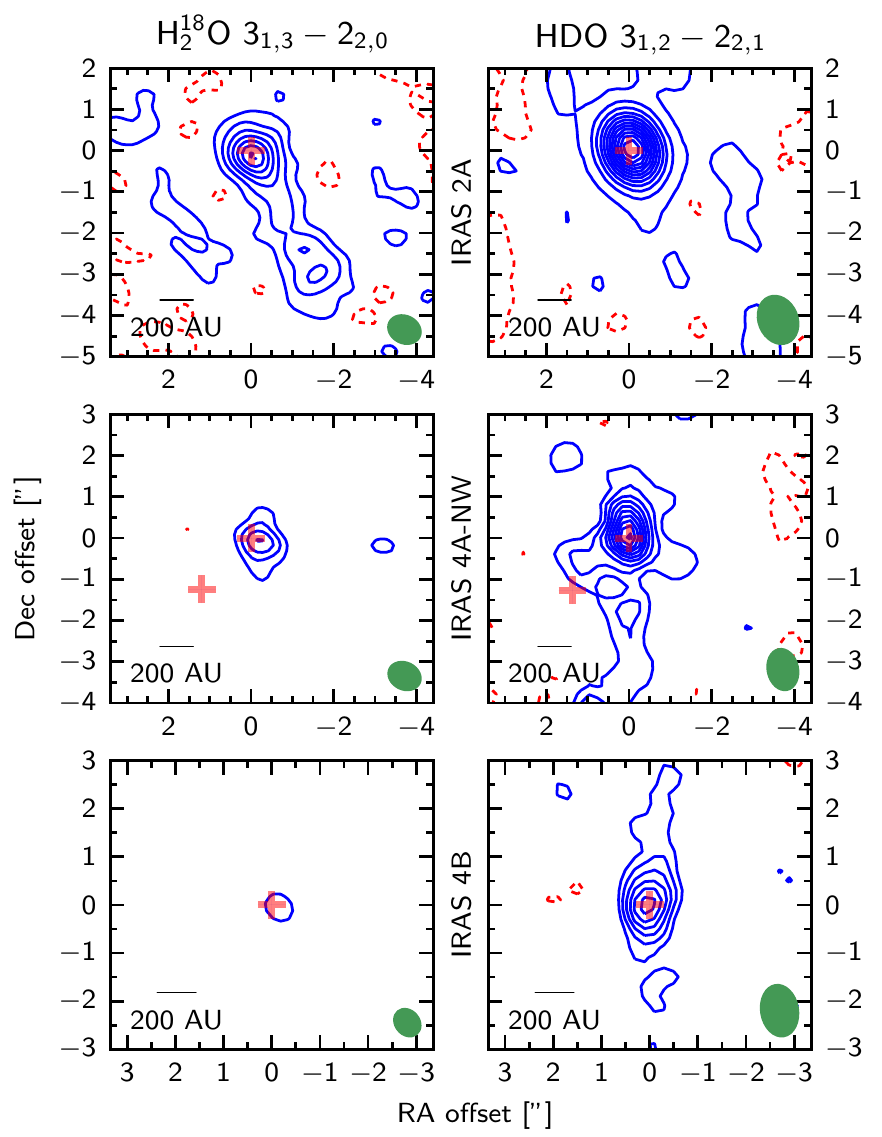}
      \caption{Integrated intensity maps for the HDO \trans{3}{1}{1}{3}{3}{1}\ and \isowater\ \trans{3}{1}{3}{2}{2}{0}\ in the interval $v_\mathrm{line}\pm$FWHM toward all three sources. The beam is shown in the lower right corner and the red crosses show the position of continuum peaks. Contours start at 41.1~m\jykms\ (5$\sigma$ steps for H$_2^{18}$O and 20$\sigma$ for HDO) for IRAS~2A, 50.1~m\jykms\ (2$\sigma$ steps) for IRAS~4A, and 29.8~m\jykms\ (2$\sigma$ steps) for IRAS~4B, dashed contours represent negative values.
              }
         \label{figure:moments}

    \includegraphics{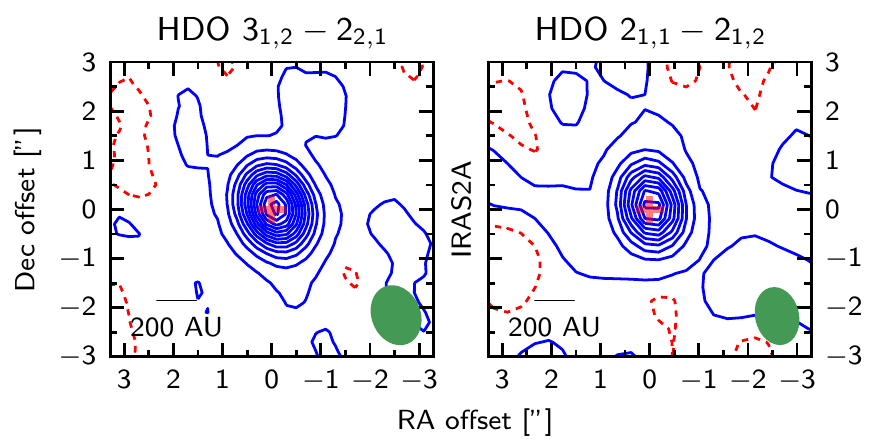}
      \caption{Integrated intensity maps for the \trans{3}{1}{2}{2}{2}{1}\ (225 GHz) and \trans{2}{1}{1}{2}{1}{2}\ (241 GHz) transitions of HDO toward IRAS~2A, in the interval $v_\mathrm{line}\pm$FWHM. The beam is shown in the lower right corner and the red crosses show the position of continuum peaks. Contours start at 39.6~m\jykms\ (steps of 20$\sigma$), dashed contours represent negative values.
              }	
         \label{figure:moments_i2a}
   \end{figure}

Fig.~\ref{figure:moments} shows the continuum subtracted intensity
maps for all three sources and water lines. For IRAS~2A
Fig.~\ref{figure:moments_i2a} shows the additional HDO line at 241.6~GHz
together with the 225.9~GHz line. The intensity maps for each source
are plotted on the same scale, i.e., the contour lines indicate the
same intensities.

Circular Gaussian profiles were fitted to the emission of each line in the (\textit{u,v})-plane after integrating over channels corresponding to $v_\mathrm{lsr}\pm$FWHM. Table~\ref{table:intensities} lists the FWHM of the 1D Gaussian fitted in the spectra, and size and intensity from fits in the (\textit{u,v})-plane of the integrated intensity for all the lines. The line widths roughly agree between the lines, being slightly wider in the HDO spectra (see Table~\ref{table:intensities}). 

   \begin{table}[ht]
      \caption[]{Results from fits to the spectra and the (\textit{u,v})-plane. The errors in the FWHM and size are from the statistical fits, while the uncertainty in intensity is assumed to be 20\%.}
         \label{table:intensities}
     $$      
         \begin{array}{p{0.14\linewidth}lll}
         \hline\hline
         \noalign{\smallskip}
            Source  &  \mathrm{Intensity} 			&  \mathrm{FWHM} 		 	 &  \mathrm{Size}\tblmark{a}	\\
            		&  \mathrm{[Jy\, km\, s^{-1}]} 	&	\mathrm{[km\,s^{-1}]}  	 &  \mathrm{[arcsec]}			\\
            \noalign{\smallskip}
            \hline
            \noalign{\smallskip}
                	& 	   			& \mathrm{HDO\ 3_{1,2}-2_{2,1}} & 	\\
            \noalign{\smallskip}
            I2A		& 3.98\pm0.80	& 4.1\pm0.1 & 0\farcs4\pm0.1	\\
            I4A-NW	& 1.97\pm0.39	& 3.7\pm0.2 & 1\farcs6\pm0.1	\\
            I4B		& 0.36\pm0.07	& 2.5\pm0.2 & 0\farcs5\pm0.1 	\\
            \noalign{\smallskip}
            \hline
            \noalign{\smallskip}
            	  	&  			& \mathrm{HDO\ 2_{1,1}-2_{1,2}}  & 				\\
           	\noalign{\smallskip}
            I2A		& 3.88\pm0.78 & 4.0\pm0.1 & 0\farcs5\pm0.1 \\
            \noalign{\smallskip}
            \hline
            \noalign{\smallskip}
                	& 	   			& \mathrm{H_2^{18}O\ 3_{1,3}-2_{2,0}} & 	\\
            \noalign{\smallskip}
            I2A		& 0.98\pm0.2	& 4.0\pm0.2 & 0\farcs8\pm0.1	\\
            I4A-NW	& 0.27\pm0.05	& 2.9\pm0.3 & 0\farcs6\pm0.1	\\
            I4B		& 0.09\pm0.02	& 0.9\pm0.1 & 0\farcs2\pm0.1 	\\
            \hline
         \end{array}
     $$ 
         \tablefoottext{a}{FWHM of circular Gaussian fit in the (\textit{u,v})-plane.}
   \end{table}


\section{Analysis}\label{section:analysis}

\subsection{Spectra and maps}

The spectra show similar characteristics for the observed sources and
lines. The emission is compact and traces the warm water within 150~AU
of the central sources. The integrated map for the IRAS~4A protobinary
shows extended emission, partly aligned with the southern,
blue-shifted outflow \citep{jorgensen07}. In contrast, the south-west
outflow of IRAS~2A seen in H$_2^{18}$O is not detected in either of
the two HDO lines. The ratio between the HDO line(s) and the
H$_2^{18}$O line is similar for IRAS~2A and IRAS~4B, with HDO showing
stronger emission than that of H$_2^{18}$O by a factor of four. In
IRAS~4A-NW, the HDO $3_{1,2}-2_{2,1}$ line is even brighter - about
seven times stronger than the H$_2^{18}$O line. Methyl formate
(CH$_3$OCHO) is clearly seen toward IRAS~4A-NW and IRAS~4B in the
225~GHz spectra, while it is absent in the spectrum toward
IRAS~2A. All lines toward IRAS~4B show very compact emission.

\subsection{Column densities, excitation temperature and the HDO/H$_2$O ratio}\label{section:column_densities}

With two transitions of HDO observed toward IRAS~2A with different
upper energy levels ($E_\mathrm{u}$), a rough estimate of the
excitation temperature ($T_\mathrm{ex}$) can be made. One issue is
that the energy levels are quite close ($167$ vs. $95$~K), which
increases the uncertainty of the determined excitation
temperature. Assuming that the HDO is optically thin and in LTE, and that the emission fills the beams
uniformly, the excitation temperature for HDO in IRAS~2A is
$T_\mathrm{ex} = 124 \pm 60$~K, using a 20\% flux calibration
uncertainty. This excitation temperature reflects the bulk temperature of the medium averaged over the entire synthesized
beam.

We can estimate the column density of HDO over the observed beam in all of the sources by assuming LTE with $T_\mathrm{ex}=124$~K as determined for HDO in IRAS~2A, that the mission fills the beam and is optically thin. For IRAS~2A this is
$11.9\pm2.4\times10^{15}~\mathrm{cm^{-2}}$, for IRAS~4A-NW it is
$8.8\pm1.8\times10^{15}~\mathrm{cm^{-2}}$, and for IRAS~4B
$1.5\pm0.3\times10^{15}~\mathrm{cm^{-2}}$. Assuming
$T_\mathrm{ex}=124$~K for the H$_2^{18}$O line as well gives column
densities for the three sources consistent with the results of
\citet{persson12}, where $T_\mathrm{ex}=170$~K was assumed.

With the column density of both HDO and H$_2^{18}$O, we deduce the
water deuterium fractionation in all three protostars. For IRAS~2A the
HDO/H$_2$O ratio is then $7.4\pm2.1\times10^{-4}$, for IRAS~4A-NW it
is $19.1\pm5.4\times10^{-4}$ and for IRAS~4B it is
$5.9\pm1.7\times10^{-4}$.  In the calculations a
\element[][16]{O}/\element[][18]{O} ratio of 560 was assumed,
appropriate for the local interstellar medium \citep{wilson94}. For
IRAS~4B, the deduced HDO/H$_2$O ratio agrees within the
uncertainties with the upper limit derived by \citet{jorgensen10b}
using lower sensitivity SMA data.

\begin{figure}[ht]
   \centering
    \includegraphics[width=\linewidth]{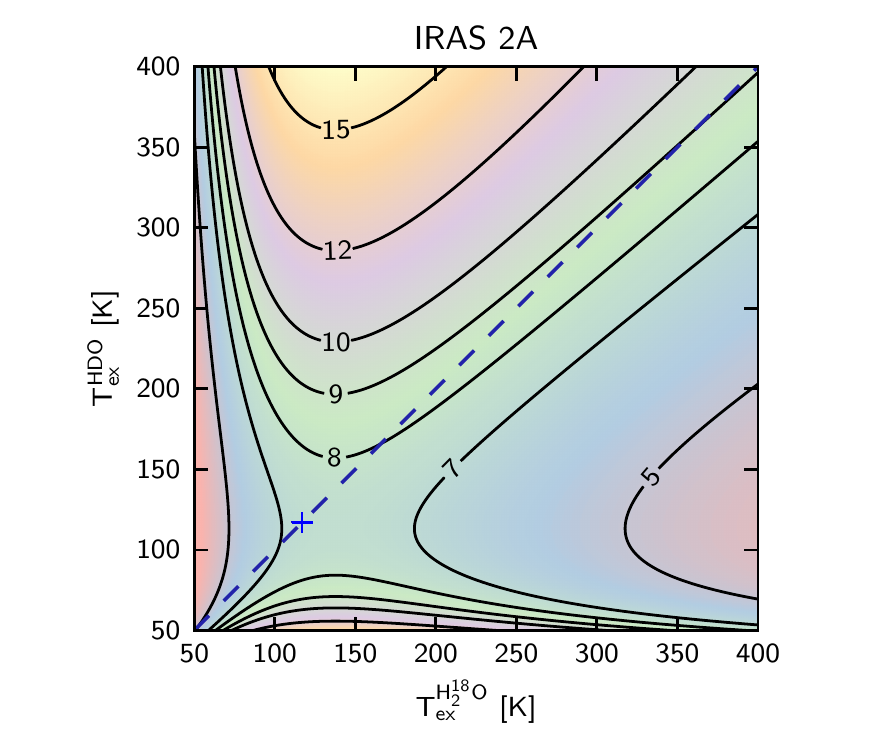}
      \caption{The variation of the water deuterium fractionation (HDO/H$_2$O) as a function of excitation temperature (40-400~K) for the observed water isotopologues. Contour values are $\times 10^{-4}$ and the dashed blue line shows where $T_\mathrm{ex}^\mathrm{H_2^{18}O} = T_\mathrm{ex}^\mathrm{HDO}$, and the cross the value determined for IRAS~2A, i.e.\ at $T_\mathrm{ex}=124$.}
         \label{figure:frac_vs_tex}
   \end{figure}

To investigate the effects of different excitation temperatures,
Fig.~\ref{figure:frac_vs_tex} shows the HDO/H$_2$O ratio in IRAS~2A in
the $T_\mathrm{ex}$ range $40-400$~K. As can be seen in the figure,
the HDO/H$_2$O ratio does not change significantly over this wide
range of excitation temperatures: even if
$T_\mathrm{ex}(\mathrm{H_2^{18}O})$ is significantly different from
$T_\mathrm{ex}(\mathrm{HDO})$, the derived abundance ratios only vary
by a factor of 3 from $5-15\times10^{-4}$.

\subsection{Radiative transfer modeling}

To test our assumption of optically thin lines used to derive column
densities in Sect.~\ref{section:column_densities}, full radiative
transfer models of spherically symmetric envelopes have been run. These
same type of models have been used to interpret the water and HDO
emission lines observed by \herschel\ and single-dish ground based
telescopes \citep{coutens12,coutens13}. In these models, the molecular
excitation is computed at each position in the envelope and subsequent
line radiative transfer is performed. The interferometric data
originate in the inner $\sim100$~AU which is the scale on which the
physical models are not well constrained. Previous observations have
shown that flattened disk-like structures appear on scales
$\lesssim300$~AU \citep{jorgensen05b,chiang12}. Because of the lack of
proper models, we adopt the spherically symmetric structure to assess the validity of the optically thin LTE approximation. Radiative transfer models were run for two cases: IRAS~16293-2422 and IRAS~2A.

The physical models start with a density structure which follows a
power-law structure, as specified below for each of the sources.  The
dust temperature is then computed self-consistently as function of
radius in the envelope for a given luminosity using the {1-D}
spherical dust continuum radiative transfer code TRANSPHERE
\citep{dullemond02}. For the typical densities of the
protostellar envelopes the gas is expected to be coupled to the dust
and we thus assume that the gas and dust temperatures are identical
\citep{jorgensen02}. The resulting temperature and density structure
are then used as input to the Monte Carlo radiative transfer code
RATRAN \citep{hogerheijde00}. Since the line profiles do not show any
complicated velocity structures the models only include a doppler
broadening parameter (the ``doppler-b parameter''
$0.6\times$FWHM). The outer radius is defined as where
$T_\mathrm{dust}=10$~K or $n=10^4$~cm$^{-3}$, whichever comes
first. HDO collisional rate coefficients of \citet{faure12} are used to calculate non-LTE population levels \citep[LAMBDA database,][]{schoier05}. To produce the collisional rate
coefficients for para-H$_2^{18}$O, the rates for H$_2^{16}$O from
\citet{daniel11} for $5<T<1500$~K were combined with the tabulated
levels for para-H$_2^{18}$O from the JPL catalog \citep{pickett98}.
The abundance of HDO or p-H$_2^{18}$O is then varied to obtain the best
fit to the observed lines.

To determine the H$_2$O abundance, the best-fit p-H$_2^{18}$O
abundance has to be multiplied with the fraction of para to total
water ratio ($4$, i.e.\ $o/p = 3/1$) and the
\element[][16]{O}/\element[][18]{O} ratio
\citep[$560$,][]{wilson94}. The molecular abundance is assumed to
follow a ``jump'' abundance profile \citep[e.g.,][]{schoier04}, with
an inner and outer abundance that changes discontinuously at the point
in the envelope where $T_\mathrm{dust}=100$~K, i.e., where water
evaporates off the dust grains \citep{fraser01}. The radiative
transfer model gives as output a model image cube with units Jy/pixel;
this is then convolved with a beam corresponding to the observations
and the integrated intensity is compared to those observed in
Table~\ref{table:intensities} and \citet{persson13a}. To check where
the emission in the relevant lines originates, the excitation is first
solved for a model that covers the entire envelope, i.e.\ out to
$r_\mathrm{out}$. Then an increasing number of cells are included in
the ray-tracer, and the resulting integrated intensities are
compared. This shows that the majority ($>97\%$) of the emission in
the observed lines comes from the region where $T\geq100$~K. To increase the numerical convergence, the envelope model for IRAS~2A was truncated at relatively small radii ($ 6100$~AU). The less dense and cold outer parts ($R>6100$~AU) of
the envelope does not contribute significantly to any absorption or emission of the observed lines.

For the IRAS~16293-2422 envelope, we adopt the same physical structure
as developed by \citet{crimier10}, which also allows us to compare our
results to \citet{coutens12}.  
In the original model a velocity field describing infall was used, here we set the turbulence doppler-b parameter to 1~\kms\ because we are only interested in fitting the total integrated intensity. Unless the
lines are significantly optically thick, the adopted doppler-b
parameter will only affect the line profile and not the total
integrated intensity. The density structure of the envelope is
described by \citep{shu77} as

\begin{eqnarray}
n(r) = n(r_0) \left( \dfrac{r}{r_0} \right)^{-1.5} \qquad r \leq r_\mathrm{inf}\\
n(r) = n(r_0) \left( \dfrac{r}{r_0} \right)^{-2} \qquad r > r_\mathrm{inf}
\end{eqnarray} 

where $r_\mathrm{inf} = 1280$~AU, $r_0=76$~AU and $n(r_0) =
2\times10^{8}$~cm$^{-3}$ (the H$_2$ number density at $r_0$ where
$T_\mathrm{dust}=100$~K).
The dust opacity as a function of the frequency is set to a power-law
emissivity model, i.e.,
\begin{equation}
\kappa = \kappa_0 \left( \dfrac{\nu}{\nu_0} \right) ^{\beta}
\end{equation}
where $\beta=1.8$, $\kappa=15$~cm$^2$\,g$_\mathrm{dust}^{-1}$ and
$\nu_0=10^{12}$~Hz. This dust opacity relation was adopted by
\citet{coutens12} in order to fit the continuum observed with the
\herschel-HIFI from 0.5 to 1 THz. The cold outer abundance at
$T<100$~K is fixed to $X_\mathrm{out} = 3 \times 10^{-11}$ as found by \citet{coutens12} for both HDO and H$_2^{18}$O with
respect to H$_2$. Changing it within a reasonable interval
($\pm10\times X_\mathrm{out}$) does not change the derived warm inner abundances
significantly.

The best fit to the 203~GHz H$_2^{18}$O line of
IRAS~16293-2422 requires an inner abundance of $X_\mathrm{in} = 4.5 \pm
0.9 \times 10^{-8}$ ($1.0 \pm 0.2 \times 10^{-4}$ for H$_2$O), and for
the HDO line at 225.9~GHz requires $X_\mathrm{in} = 11.0 \pm 2.2 \times
10^{-8}$. These abundances result in a HDO/H$_2$O ratio of $10.9 \pm
3.1 \times 10^{-4}$. The models show that both HDO lines at 225.9~GHz
and 241.6~GHz are marginally optically thick whereas the H$_2^{18}$O
line is neither masing nor optically thick. The derived abundance ratio
clearly agrees with the previously stated LTE, optically thin value of
$9.2\pm 2.6\times10^{-4}$ within the uncertainties \citep{persson13a}. 
If the o/p ratio for the main collision partner (H$_2$) is changed to 0, the HDO/H$_2$O ratio goes down by 20\%. For water, if an o/p ratio (H$_2$O) of 1 and a $^{16}$O/$^{18}$O ratio of 400 is assumed the HDO/H$_2$O ratio increases by a factor of 2.8. This shows that the assumed o/p ratio of either H$_2$O or H$_2$ and the $^{16}$O/$^{18}$O ratio can affect the resulting HDO/H$_2$O ratio down by 20\% and up by a factor of 2.8.

For IRAS~2A we have used the model described by \citet{kristensen12}
and it extends from $35.9$~AU (terminates where $T>250$~K) to
$17950$~AU (but truncated at 6100~AU). The density profile is a single
power-law, $n(r) = n(r_0)\,(r/r_0)^{-1.7}$, where $r_0 =
r_\mathrm{in}$ and $n(r_0) = 4.9\times10^{8}$~cm$^{-3}$. The dust
opacity is set to Table~1, Column~5 in \citet{ossenkopf94} (e.g., OH5
- dust grains with thin ice mantles) and the dust temperature is again
computed self-consistently. The outer abundance is fixed to
$X_\mathrm{out} = 3 \times 10^{-11}$ for both HDO and H$_2^{18}$O, and
as for IRAS 16293-2422, changing it within a reasonable interval
($\pm10\times$) does not change the derived abundances
significantly. The best fit inner abundance for the H$_2^{18}$O line
is $X_\mathrm{in} = 4.8 \pm 1.0 \times 10^{-8}$ ($1.1 \pm 0.2 \times
10^{-4}$ for H$_2$O), and $X_\mathrm{in} = 34.0 \pm 6.8 \times
10^{-8}$ for the HDO line at 225.9~GHz. This corresponds to a
HDO/H$_2$O ratio of $31.6 \pm 8.9 \times 10^{-4}$. While the HDO
abundance derived from the 225~GHz line also fits the 241~GHz line within the uncertainties, the 143~GHz line observed by \citet{taquet13b} is fit by a somewhat lower HDO abundance of $X_\mathrm{in}=24.3\pm4.9$ giving a HDO/H$_2$O ratio of $22.3\pm6.4\times10^{-4}$. All these HDO lines give a HDO/H$_2$O ratio that is
higher by a factor of $3-4$ than the optically thin LTE value of $7.4\pm 2.1
\times 10^{-4}$. This difference is largely due to the optical depth effects of the HDO lines.

The models for both IRAS 16293-2422 and IRAS 2A indicate that the HDO
lines are marginally optically thick. This could indicate that the
derived HDO column densities from the optically thin calculation (in
Section~\ref{section:column_densities}) are slightly underestimated
for all four sources. On the other hand, the H$_2^{18}$O line at
203.4~GHz is well constrained by LTE and is optically thin.  Combined,
this would imply an underestimate of the HDO/H$_2$O ratio. However,
the models show that this effect is minor, a factor of $\sim3$ or less based on a comparison of the ratios derived from the radiative transfer models and the direct LTE estimates. The IRAS
16293-2422 example also shows that differential excitation effects of
HDO versus H$_2^{18}$O can counteract the HDO optical depth effects.

\section{Discussion and Conclusions}\label{section:discussion}

All deduced HDO/H$_2$O ratios are consistent with our previous low
estimates for two deeply-embedded protostars \citep{persson13a,jorgensen10b}. The consistently low HDO/H$_2$O ratios for all four sources suggests that the chemical evolution of water for these sources is similar.

Our high-resolution observations make it possible to focus on the
compact warm gas.  The sources show some extended emission in HDO
and/or H$_2^{18}$O that would be included in any lower-resolution
(single-dish) observations. Another key to our conclusions is that the
observed lines are optically thin or at most marginally optically
thick.  Because of their higher Einstein A coefficients, the
$3_{1,2}-3_{0,3}$ lines of H$_2^{16}$O and H$_2^{18}$O observed with
\herschel\ rapidly become optically thick, each at slightly different
radii complicating the analysis \citep{visser13}. Unless the optical
depth is properly treated, the HDO/H$_2$O ratios deduced using
\herschel\ data are likely to be overestimated. The ratio estimated for IRAS~2A by \citeauthor{visser13} accounting
for the optical thickness of the \herschel\ lines gives a HDO/H$_2$O
ratio of $1\times10^{-3}$, in agreement with our estimates.

The recently deduced warm HDO/H$_2$O ratios for IRAS~2A
($3\times10^{-3}$) and IRAS~4A ($5\times10^{-3}$) at the relevant
densities ($n_\mathrm{H_2}=1\times10^{8}$~cm$^{-3}$) by
\citet{taquet13b} are factors of 3--5 higher than our estimates. With a
lower spectral ($4$\ vs. $0.1$~\kms) and spatial
(2\arcsec\ vs. 0.8\arcsec) resolution than that of our H$_2^{18}$O
data, the HDO observations used by \citeauthor{taquet13b} probe larger
scales, thus are affected by extended emissions as observed in our observations at scales $\gtrsim500$~AU (with higher HDO/H$_2$O ratios as a probable result). 
Furthermore, the upper energy levels of the HDO and H$_2^{18}$O transitions differ
significantly, $E_\mathrm{u}=319$~K for the HDO $4_{2,2} - 4_{2,3}$
transition vs.\ 204~K for the H$_2^{18}$O $3_{1,3} - 2_{2,0}$. The
transitions could originate in slightly different regions depending on
the source structure.

\citet{coutens13} used \herschel\ and several ground-based single dish
telescopes to observe lines of HDO toward IRAS~4A and IRAS~4B. From 1D
radiative transfer models and using the H$_2^{18}$O column density by
\citet{persson12}, they estimated HDO/H$_2$O ratios in the inner warm
regions of IRAS~4A ($4.0-30.0\times10^{-4}$) and IRAS~4B
($1.0-37.0\pm10^{-4}$). To reproduce the absorption seen in some of
the spectral lines, a cold foreground absorbing layer is
introduced. The results for the warm HDO/H$_2$O ratio show a wide range but are consistent with our findings, namely that the ratio is
similar for the two protostars, and that it is at most a few
$\times10^{-3}$.

The radiative transfer modeling highlights the uncertainties
involved in the derivation of the HDO/H$_2$O ratio. However, the
ratios deduced with the optically thin LTE model-independent (i.e.,
not employing physical models and radiative transfer calculations to
interpret the emission) method used in this paper and in
\citet{persson13a} is direct, and agrees with the full modeling within
a factor of $3-4$. Until the physical structure on small scales can be
constrained, running full radiative transfer models can not yield a
more accurate HDO/H$_2$O ratio in the warm gas. However, spherically symmetric radiative
transfer modeling can give indications on non-LTE effects and is a
valid approximation for the outer envelope
\citep[e.g.,][]{mottram13}. Furthermore, the modeling also confirms
that the observed transitions of the water isotopologues are not
masing in these environments.

The emerging picture of the water deuterium fractionation in warm
inner regions of deeply-embedded low-mass protostars is therefore that
of a consistent and relatively low ratio. While the protostars in this
study are all located in the same star forming cloud, NGC~1333,
IRAS~16293-2422 in \citet{persson13a} is located in a different cloud,
$\rho$~Ophiuchus. Its similar HDO/H$_2$O ratio indicates that a
different birth environment does not necessarily imply a different
ratio. To confirm this, HDO/H$_2$O ratios determination toward more sources are needed.

The infall timescales for the early, deeply-embedded stages are
generally too short for any significant gas-phase chemical processing
of water to take place in the warm $>$100 K gas before the material
enters the disk \citep{schoier02}. Therefore the gaseous water
observed here probably originates by sublimation of ice formed earlier
in the evolution.

\begin{figure*}[ht]
   \centering
    \includegraphics{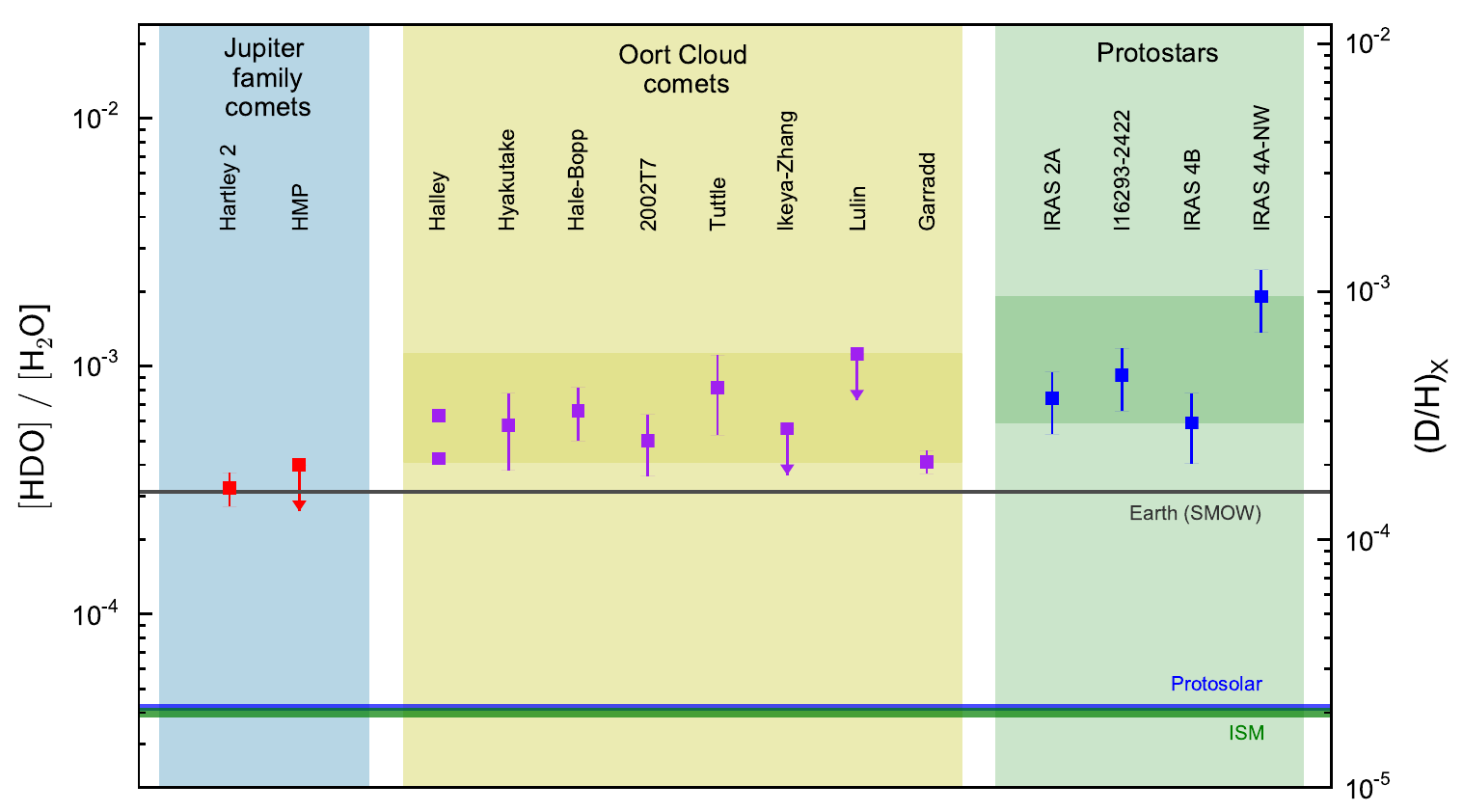}
    \caption{Values of the HDO/H$_2$O ratio, for different
      objects. For references, see Table~\ref{table:ratios}. Error bars
      for protostars reflect those obtained using the optically thin
      LTE approximation.  }
         \label{figure:dhfractionation}
   \end{figure*}

In Fig.~\ref{figure:dhfractionation} the deuterium fractionation of
water for all four low-mass protostars (IRAS~2A, IRAS~4A-NW, IRAS~4B
and IRAS~16293-2422) is presented together with those found in comets,
Earth, interstellar medium (ISM) and protosolar values. The small
differences in the amount of deuteration seen toward the inner regions
of protostars and the comets in our solar system suggests that only
small amounts of processing of water takes place between the deeply
embedded stages and the formation of comets and planetesimals. If
water was delivered to Earth by comets and small solar system bodies
during the early evolution of the Sun, this would be the same water
that was formed directly on the grain surface in the cold early stages
of the formation of our Sun.

\section{Summary and Outlook}
In this paper high-resolution interferometric observations of the HDO
$3_{1,2}-2_{2,1}$ line at 225.9~GHz toward three deeply embedded
protostars have been presented. In addition, the HDO $2_{1,1}-2_{1,2}$
line was observed toward one of the sources. With previous
observations of H$_2^{18}$O \citep{persson12}, we deduce a model independent HDO/H$_2$O ratio.

\begin{itemize}

\item Observations of two HDO lines at high angular resolution give an
  excitation temperature of $T_\mathrm{ex}=124 \pm 60$~K for IRAS~2A.

\item Assuming that the water emission is optically thin and is in LTE
  at $T_\mathrm{ex}=124$~K for all three sources, we derive a
  HDO/H$_2$O ratio of $7.4\pm2.1\times10^{-4}$ for
  IRAS~2A, $19.1\pm5.4\times10^{-4}$ for IRAS~4A-NW
  and $5.9\pm1.7\times10^{-4}$ for IRAS~4B.

\item The high-resolution interferometric observations give HDO/H$_2$O
  ratios in a model independent way with an accuracy of a factor of
  3--4, as confirmed by using non-LTE radiative transfer models. The fact that
  the  observed lines have much lower optical depths than those
  observed with \herschel\ explains much of the early discrepancies in
  inferred values.

\item The HDO/H$_2$O ratios deduced for these deeply-embedded
  protostars using interferometric observations are consistent from
  source to source. The ratios range from $5.9$ to $19.1\times10^{-4}$, close to those observed
  in solar system comets originating in the Oort cloud. The small
  difference between these reservoirs indicates little processing of
  material has taken place between the deeply-embedded, collapsing
  stage and the formation of comets/planetesimals.

\end{itemize}

\begin{acknowledgements}
We wish to thank the IRAM staff, in particular Arancha Castro-Carrizo
and Chin Shin Chang, for their help with the observations and
reduction of the data.  We also appreciate discussions with Joe
Mottram and Audrey Coutens on various aspects of water modeling. IRAM
is supported by INSU/CNBRS(France), MPG (Germany), and IGN (Spain). The
research at Centre for Star and Planet Formation is supported by the
Danish National Research Foundation and the University of
Copenhagen\rq{s} programme of excellence. This research was
also supported by a Junior Group Leader Fellowship from the
Lundbeck Foundation to JKJ.  EvD acknowledges the Netherlands
Organization for Scientific Research (NWO) grant 614.001.008. EvD and
MVP acknowledges EU FP7 grant 291141 CHEMPLAN. This work has benefited
from research funding from the European Community's sixth Framework
Programme under RadioNet R113CT 2003 5058187.
\end{acknowledgements}

\bibliography{bibfile}

\begin{appendix}

\section{Tables}

\begin{table*}[h]
\begin{centering}
    \caption{The HDO/H$_2$O abundance ratios for different objects. OCC - Oort Cloud Comet, JFC - Jupiter Family Comet. Some values are $2\times$ the D/H ratio and that the ratio in protostars refers to the warm gas in the inner regions.}
    \label{table:ratios}
\begin{tabular}{lllllr}
\hline\hline
\noalign{\smallskip}
Type 			& Object 				& Origin	  		& HDO/H$_2$O 		 & Tracers	& Ref.  \\ 
	 			& 						&  	  				&  	[$\times10^{-4}$]& (Proxy)	&   		 \\ 
\noalign{\smallskip}
\hline
\noalign{\smallskip}
	Comet  		& 1/P Halley 			& (OCC)				& $6.3\pm0.7$ 		& H$_3$O$^{+}$, H$_2$DO$^{+}$	&    1 \\ 
	Comet  		& C/1996 B2 Hyakutake 	& (OCC) 		  	& $5.8\pm2.0$		& H$_2^{16}$O, HDO &    2 \\ 
	Comet  		& C/1995 O1 Hale-Bopp 	& (OCC)			 	& $6.6\pm1.6$		& HDO, OH &    3 \\ 
	Comet  		& 8P/Tuttle 			& (OCC) 			& $8.2\pm2.9$		& H$_2^{16}$O, HDO &    4 \\ 
	Comet  		& C/2007 N3 Lulin 		& (OCC)				& $<11.2$			& H$_2^{16}$O, HDO &    5 \\ 
	Comet  		& C/2002 T7 LINEAR 		& (OCC)				& $5\pm1.4$\tblmark{a} & OH, $^{18}$OH, OD			&    6 \\ 
	Comet  		& 153P Ikeya-Zhang 		& (OCC)				& $<5.6\pm0.6$		& H$_2^{16}$O, H$_2^{18}$O, HDO		&    7 \\ 
	Comet  		& C/2009 P1 Garradd 	& (OCC)				& $2.06\pm0.22$		& H$_2^{18}$O, H$_2$O, HDO &    8 \\ 
	\noalign{\vskip1pt}
	\hline
	\noalign{\smallskip}
	 &  OCC Mean\tblmark{c}		&     				& $6.4\pm1.0$		& 			&      \\ 
\hline
\noalign{\smallskip}

	Comet  		& 103P Hartley 2 		& (JFC)				& $3.2\pm0.5$		& H$_2^{18}$O, HDO &    9 \\ 
	Comet  		& 45P Honda-Mrkos-Pajdusakov (HMP) 				& (JFC)				& $<4.0$				& H$_2^{16}$O, H$_2^{18}$O, HDO &    10 \\ 

\noalign{\smallskip}
\hline
\noalign{\smallskip}

	Planet  	& Earth (SMOW)	  		&  					& $3.114\pm0.002$ 	& 			&   11 \\ 

\noalign{\smallskip}
\hline
\noalign{\smallskip}
				& Protosolar  			&  					& $0.42\pm0.08$\tblmark{b}	& $^3$He, $^4$He & 	12 \\ 
				& (L)ISM\tblmark{c} 	&  					& $\geq0.40\pm0.02$			& H, D & 	13 \\ 

\noalign{\smallskip}
\hline
\noalign{\smallskip}
	Protostar  	& IRAS 16293-2422 		& 	$<50$~AU		& $9.2\pm2.6$  		& H$_2^{18}$O, HDO 	&  14  \\ 
	Protostar  	& 				 		& 					& $340$				& H$_2^{17}$O, H$_2^{18}$O, HD$^{18}$O, HDO 	& 15 \\
	Protostar  	& 				 		& 					& $300$				& H$_2^{16}$O, HDO 		& 16   \\  
	Protostar  	& NGC 1333-IRAS~2A 		&  $\lesssim300$~AU	& $7.4\pm2.1$  		& H$_2^{18}$O, HDO 	& 17 \\ 
	Protostar  	& 						& 					& $\geq100$  		& H$_2^{16}$O, HDO 	& 18 \\
	Protostar  	& 						& 					& $30-800$  		& H$_2^{18}$O, HDO &  19 \\
	Protostar  	& NGC 1333-IRAS~4A-NW	& $\lesssim300$~AU	& $19.1\pm5.4$		& H$_2^{18}$O, HDO 	& 17 \\ 
	Protostar  	& 				 		& 					& $50-300$  		& H$_2^{18}$O, HDO & 19   \\ 
	Protostar  	& 				 		& 					& $4-30$  			& H$_2^{18}$O, HDO & 20   \\ 
	Protostar  	& NGC 1333-IRAS~4B 		& $\lesssim300$~AU	& $5.9\pm1.7$		& H$_2^{18}$O, HDO 	& 17 \\
	Protostar  	& 				 		& 					& $1-37$  			& H$_2^{18}$O, HDO 	& 20  \\ 
	Protostar  	& 				 		& 					& $<6.4$  			& H$_2^{18}$O, HDO 	& 21  \\ 
\noalign{\smallskip}
\hline   
\end{tabular}\\
        \tablefoottext{a}{Value is $2\times$D/H.}
		\tablefoottext{b}{No upper limits included}
        \tablefoottext{c}{Local Interstellar Medium, i.e. $\sim1-2$~kpc from the Sun}
  (\textbf{1}) \citet{eberhardt95}; 
  (\textbf{2}) \citet{bockelee98};
  (\textbf{3}) \citet{meier98};
  (\textbf{4}) \citet{villanueva09};
  (\textbf{5}) \citet{gibb12};
  (\textbf{6}) \citet{hutsemekers08};
  (\textbf{7}) \citet{biver06};
  (\textbf{8}) \citet{bockelee12};
  (\textbf{9}) \citet{hartogh11};
  (\textbf{9}) \citet{lis13};
 (\textbf{11}) \citet[][and ref. therein]{delaeter03};
 (\textbf{12}) \citet{geiss98,lellouch01};
 (\textbf{13}) \citet[][and ref. therein]{prodanovic09};
 (\textbf{14}) \citet{persson13a};
 (\textbf{15}) \citet{coutens12};
 (\textbf{16}) \citet{parise05};
 (\textbf{17}) This work;
 (\textbf{18}) \citet{liu11};
 (\textbf{19}) \citet{taquet13b}
 (\textbf{20}) \citep{coutens13}
 (\textbf{21}) \citet{jorgensen10b}
\end{centering}
\end{table*}

\begin{table}[ht]
    \caption{\label{table:molecules} Molecular data for the different
      lines, from JPL \citep{pickett98} and CDMS
      \citep{muller01}. }

\begin{tabular}{lllllllll}
\hline\hline
\noalign{\smallskip}                                                                             
 Molecule	& Transition				& Rest Freq.	& $E_\mathrm{u}$	& Strength\\ 
         	&							& [GHz]			& [K]				& [$\log_{10}{A_{ul}}$]\\
\noalign{\smallskip}
\hline                                                                                                 
\noalign{\smallskip}
\isowater	&  \trans{3}{1}{3}{2}{2}{0}	& 203.40752		& 203.7				& -5.3177\\
HDO			&  \trans{3}{1}{2}{2}{2}{1}	& 225.89672		& 167.6				& -4.8799\\
HDO			&  \trans{2}{1}{1}{2}{1}{2}	& 241.56155		& 95.2				& -4.9256\\
\noalign{\smallskip}
\hline
\end{tabular}
\end{table}

\end{appendix}

\end{document}